\documentclass[aps,prx,twocolumn,floatfix,longbibliography,nofootinbib]{revtex4-2}
\usepackage[utf8]{inputenc}
\usepackage{natbib}
\usepackage{graphicx}
\usepackage{xcolor}
\usepackage[tbtags]{amsmath}
\usepackage[colorlinks, linkcolor=blue]{hyperref}
\hypersetup{colorlinks,allcolors=black}
\usepackage{amssymb}
\usepackage{gensymb}
\usepackage{txfonts}
\usepackage{comment}
\usepackage{float}
\usepackage{amsmath}
\usepackage{tabularx,graphicx}
\usepackage{epstopdf}
\usepackage{latexsym}
\usepackage{color, colortbl}
\usepackage{psfrag}
\usepackage{bbm}
\usepackage{bm}
\usepackage{titlesec}
\usepackage{dsfont}
\usepackage{feynmp}
\usepackage{slashed}
\usepackage{multirow}
\usepackage[normalem]{ulem}
\renewcommand{\vec}[1]{\boldsymbol{#1}}

\def \k {{\vec k}}

\def \beq {\begin{eqnarray}}
\def \eeq {\end{eqnarray}}
\def \tn {\textnormal}

\def \nn {\nonumber}

\begin{document}
\title{Correlated Topological Mixed-Valence Insulators in  Moir\'e Hetero-Bilayers} 
\author{Juan Felipe Mendez-Valderrama}\thanks{These authors contributed equally to this work}
\author{Sunghoon Kim}\thanks{These authors contributed equally to this work}
\author{Debanjan Chowdhury}
\affiliation{Department of Physics, Cornell University, Ithaca, New York 14853, USA.}
%\email{debanjanchowdhury@cornell.edu}

\begin{abstract}
Moir\'e transition metal dichalcogenide (TMD) materials provide an ideal playground for studying the combined interplay of strong interactions and band-topology over a range of electronic fillings. Here we investigate the panoply of interaction-induced electronic phases that arise at a total commensurate filling of $\nu_T=2$ in moir\'e TMD heterobilayers, focusing specifically on their renormalized band-topology. We carry out a comprehensive self-consistent parton mean-field analysis on an interacting mixed-valence Hamiltonian describing AB-stacked MoTe$_2$/WSe$_2$ to highlight different ingredients that arise due to ``Mottness", band-flattening, an enhanced excitonic tendency, and band-inversion, leading to correlated topological semi-metals and insulators. We also propose a possible route towards realizing fractionalized insulators with emergent neutral fermionic excitations in this and other closely related platforms.
\end{abstract}

\maketitle

{\it Introduction.-} Moir\'e transition metal dichalcogenide (TMD) materials have brought together the fascinating interplay of tunable electronic correlations, geometric frustration and band-topology in a highly controllable setting \cite{Wu_hetero_PRL, Wu_Macdonald_homo_prl,Mak_Shan_review}. Using dual-gated untwisted heterobilayer and twisted homobilayer TMD devices, a series of remarkable experiments have already demonstrated the emergence of local moments in a Mott insulator \cite{Tang_simulation_2020}, a sequence of Wigner-Mott insulators at rational fillings \cite{Tang_simulation_2020,Regan_mottwigner_2020,Xu_mottwigner_2020,Li_imagewigner_2021}, bandwidth-tuned continuous metal-insulator transitions at fixed fillings \cite{Ghiotto_2021,Li_2021}, and the emergence of Kondo-screening \cite{Zhao_HFL_2023} as well as tunable Kondo-breakdown \cite{Zhao_Kondobreak_2023}. On the topological front, the discovery of fractional quantum anomalous Hall insulators \cite{Cai_tMoTe2_2023,Zeng_tMoTe2_2023,Park_tMoTe2_2023,Xu_tMoTe2_PRX}, as well as fractional quantum spin Hall insulators \cite{Kang_fqsh_2024} in the same setting has really helped establish this material platform as a versatile quantum simulator of correlated many-body phenomena. This manuscript is geared towards investigating the possible emergence of a non-trivial and ``interaction-renormalized'' band-topology as a result of hybridizing relatively dispersive bands with highly correlated and nearly flat bands at commensurate fillings \cite{Faiprivate}. The primary goal here is to analyze the stability of topological ``mixed-valence" (and Kondo) insulating phases in moir\'e TMD heterobilayers, and investigate the quantum phase transitions to proximate phases.

 The study of correlated mixed-valence materials in three-dimensional compounds with strong valence (charge) fluctuations has a rich history \cite{MottMV,varma_rmp_1976,rise1,rise2}. Recent years have re-energized the field as a result of experiments in a number of strongly spin-orbit coupled compounds such as SmB$_6$, YbB$_{12}$, and FeSb$_2$, finding evidence for a high-temperature metallic state evolving into a paramagnetic charge-insulator, with a metallic surface-state at low temperature. First-principles based theory suggests that these insulators are topological \cite{DaiPRL,DaiYb,ZXFe,colemanAR}, and there is already compelling experimental indication that the surface states in SmB$_6$ \cite{smb6_1,smb6_2,smb6_3,smb6_4,smb6_5,smb6_6} as well as YbB$_{12}$ \cite{Ybb12_1,Ybb12_2} are indeed topological. However, the insulating bulk in SmB$_6$ and YbB$_{12}$ shows striking quantum oscillations \cite{Suchitra15,Sebastian18,LuLi18}; remarkably, the electrically insulating bulk in YbB$_{12}$ also acts as a metallic-like thermal conductor, violating the Wiedemann-Franz ratio by orders of magnitude \cite{Sato2019}. These observations suggest the intriguing possibility of their insulating bulk hosting emergent, electrically neutral fermionic excitations forming a three-dimensional Fermi surface \cite{baskaran,DCNC,colemanskyrme,varmamajorana}. The possible origin for {\it some} of these observations due to impurities and more conventional scenarios has been discussed \cite{Knolle1,fawang,FuQO,Knolle2,randeria,skinner}, but a conclusive picture is yet to emerge. The microscopic complexity associated with the traditional rare-earth-based compounds, and the inability to tune across the interaction-induced insulating phases readily has remained a roadblock towards resolving the key outstanding challenges. Finding alternative ways to recreate similar phenomenology in relatively simpler and tunable platforms such as the moir\'e bilayer TMD, and probe their low-energy excitations using novel methods will be an important milestone.

{\it TMD heterobilayers.-} We focus here on AB-stacked MoTe$_2$/WSe$_2$, where the lattice mismatch  gives rise to a moir\'e superlattice with lattice constant $a_{\rm M}\sim 5$ nm. The wave functions of the topmost valence bands are centered at the MM (M$\equiv$Metal) and XX (X$\equiv$Chalcogen) sites of the AB-stacking, respectively, which constitute the two distinct sublattice sites in a low-energy description in terms of a honeycomb lattice. The interlayer tunneling (referred to as the ``hybridization'' henceforth) can be treated as a perturbation relative to the decoupled layers due to the combined effects of a strong spin-momentum locking and the AB-stacking. Additionally, the MoTe$_2$ layer hosts a narrower bandwidth, and is more ``correlated'' relative to the WSe$_2$ layer; the former is the analog of the $f-$electron while the latter represents the $c-$electron in the usual scheme of describing heavy fermion materials. We will demonstrate that the bilayer is an ideal platform to realize a topological mixed-valence insulator over a wide range of parameters, but many of these considerations can be generalized to other layered setups. 

\begin{figure*}[pt!] 
\centering
\includegraphics[width=1\linewidth]{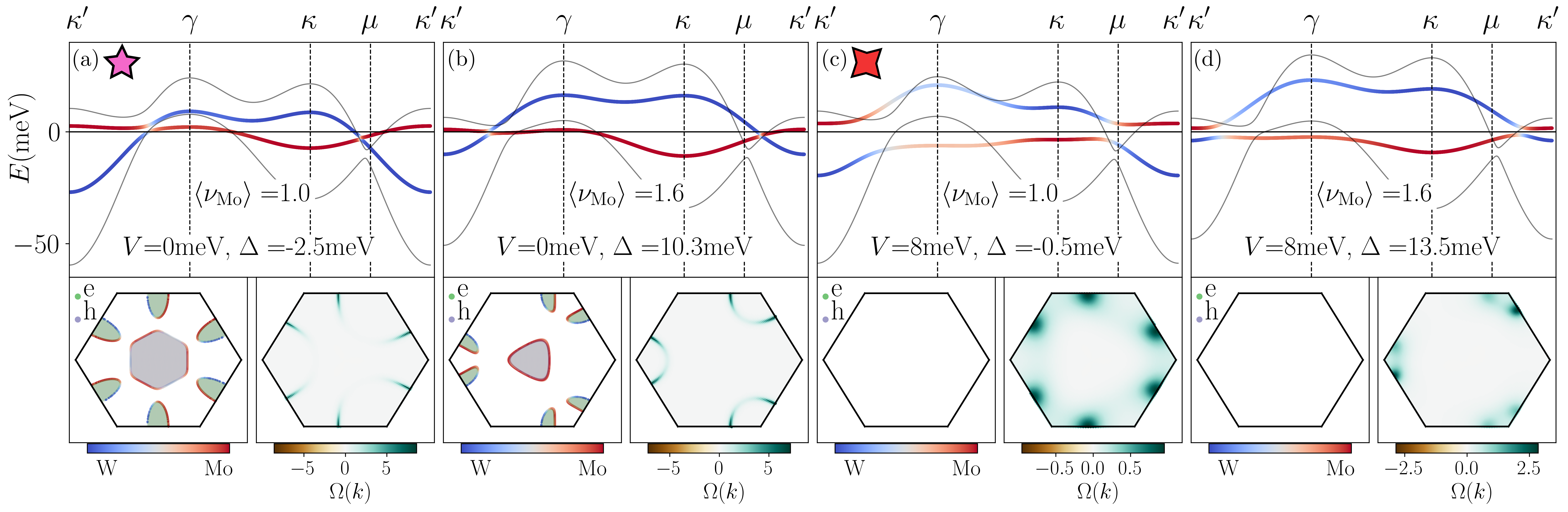}
\caption{Top: (Renormalized) electronic dispersions for valley $K$ with $U=30$ meV and other parameters as introduced in main text. Bottom: Electronic fermiology and Berry curvature distribution, $\Omega(k)$, for the lower bands. (a) and (b) show results for $V=0$, which realizes a TSM. (c) and (d) show results for $V=8$ meV, which realize a TMVI. The non-interacting bands are shown in grey in the top panel and the renormalized bands in blue/red. The representative values of $\Delta$ are chosen (see Fig.~\ref{fig:mf_pd}) to realize fillings at and away from $\langle \nu_{\tn{Mo}}\rangle\approx1$.}
\label{fig:effect_of_V}
\end{figure*} 

Previous experimental work in this setup has already revealed a topological phase transition driven by the displacement field from a non-topological Mott insulator (where only the MoTe$_2$ layer is occupied) to a quantum anomalous Hall insulator at fixed total moir\'e unit cell filling $\nu_T=1$ \cite{Li_qah_2021, Tao_PRX_valley}. A series of subsequent theoretical works have investigated this physics  \cite{Xie_valley_pol_QAH_2022,QAH_DFT_zhang_2021,QAH_devakul_2022,HPan_topo_ab_2022,Su_topo_flat,Chang_qah_prb,Rademaker_hetero_2022,Excitonic_chern_dong_zhang_2023,xie2022_valley_coherent,Xie_Pan_SDS_nematic_exciton}. Doping away from this commensurate limit, as itinerant holes are introduced to the WSe$_2$ layer while the MoTe$_2$ layer remains a Mott insulator, over a wide range of fillings an effective Kondo lattice model \cite{Guerci_2023} can be simulated \cite{Zhao_HFL_2023,Zhao_Kondobreak_2023}. Finally, at the commensurate total filling $\nu_T=2$, a displacement-field induced topological phase transition from a trivial band insulator to a ${\mathbb{Z}}_{2}$ topological insulator has been observed \cite{Zhao_qsh_2024}. Over the observed field range, the electrons mostly reside in the MoTe$_2$ layer, away from any regime of ``Mottness"; it is unclear the extent to which the strong correlation effects associated with a correlated half-filled band need to be taken into consideration \cite{HPan_topo_ab_2022}. This manuscript is devoted to the study of the insulating and semi-metallic phases, both with and without non-trivial topology, that emerge at $\nu_T=2$ taking into account the Mott-like correlations from the outset.

{\it Model.-} Instead of working in the Kondo-limit directly with $\langle \nu_{\tn{Mo}}\rangle = 1$, and treating the degrees of freedom in the MoTe$_2$ layer as $S=1/2$ local moments \cite{Guerci_2023}, we work with the full low-energy lattice model including the charge fluctuations \cite{Si24}. The model is given by $H_{\tn{MV}}= H_{\rm{Mo}} + H_{\rm{W}} + H_{\rm{Mo-W}}$ with $\ell=\{\rm{Mo}, \rm{W}\}$,
\begin{subequations}
\beq
H_{\ell} &=& -t_{\ell}\sum_{\sigma,\langle\langle i,j\rangle\rangle \in \ell } (e^{-i \nu_{ij} \phi_{\ell}s_\sigma} c^\dagger_{i\sigma}c_{j\sigma} + \tn{h.c.}) - \mu  N_{\ell} \nn\\
&& -\tau_{\ell} \frac{\Delta}{2} N_{\ell} + U_{\ell}\sum_{i\in \ell} n_{i\uparrow} n_{i\downarrow}, \\
H_{\rm{Mo-W}} &=& - t_{\rm{hyb}} \sum_{\langle i,j \rangle} (c^\dagger_{i\sigma}c_{j\sigma} + \tn{h.c.}) + V\sum_{\langle i,j\rangle} n_{i} n_{j}.
\eeq  
\end{subequations}
Here $\nu_{ij}=-\nu_{ji}$ represents the direction-dependent factor for the path connecting sites $i$ and $j$ that controls the pseudospin-dependent next-nearest neighbor hopping with an additional sign $s_\sigma$, which ensures that time-reversal symmetry is preserved between the microscopic valleys labeled by $\sigma$ \cite{QAH_DFT_zhang_2021,QAH_devakul_2022,Rademaker_hetero_2022}. The non-interacting estimates for the hopping amplitudes from DFT are $t_{\rm{Mo}}=4.5~$meV, $t_{\rm{W}}=9~$meV \cite{QAH_DFT_zhang_2021,QAH_devakul_2022}, while a simple estimate of the onsite interaction is $U\sim e^2 /\epsilon \epsilon_0 a_M$ that can range from $U\sim30-60~$meV for dielectric constants $\epsilon\sim10-5$. Due to lattice relaxation, different spin flavors can mix within a single valley leading to a weak interlayer tunneling that translates to a finite hybridization, $t_{\rm{hyb}}=2~$meV \cite{QAH_DFT_zhang_2021}. In momentum space, the hybridization form-factor reads $\Phi_\k = t_{\rm{hyb}} \sum_{j=1}^3 \tn{exp}[i\k\cdot\vec{\delta}_j]$ with $\vec\delta_j=a_{\rm M}/\sqrt{3}(\sin(2\pi j/3),-\cos(2\pi j/3))$; the form-factor plays a crucial role in the development of a topological gap. The finite $t_{\rm{hyb}}$ leads to a single $U(1)$ conserved charge associated with the total density, where the chemical potential, $\mu$, couples to $N_{\rm{Mo}}+N_{\rm{W}}$. The effect of the displacement field is incorporated via a sublattice potential, $\Delta$, that couples to both layers with opposite signs $\tau_{\rm Mo}=-\tau_{\rm W}=1$, respectively.

The $C_3$ eigenvalues of the orbitals within DFT dictate that the MoTe$_2$ and WSe$_2$ wannier functions pick up a phase of $\phi_{\rm Mo} \approx \phi_{\rm W} \approx 2\pi/3$ \cite{LiFW2021, Rademaker_hetero_2022, QAH_DFT_zhang_2021}. 
Consequently, the band maxima of both the MoTe$_2$ and WSe$_2$ bands reside near the $\kappa$ and $\kappa^{\prime}$ points of the moir\'e Brillouin zone. These $C_3$ eigenvalues and the associated location of the band maxima are crucially important for determining the band-topology as a function of $\Delta$ for a finite $t_{\rm{hyb}}$ \cite{QAH_devakul_2022}. Additionally, the nearest-neighbor (interlayer) Coulomb interaction, $V$, can lead to qualitatively new effects and is not expected to be small from a purely microscopic perspective \cite{Excitonic_chern_dong_zhang_2023}. 
 
\begin{figure*}[pth!]
\centering
\includegraphics[width=1\linewidth]{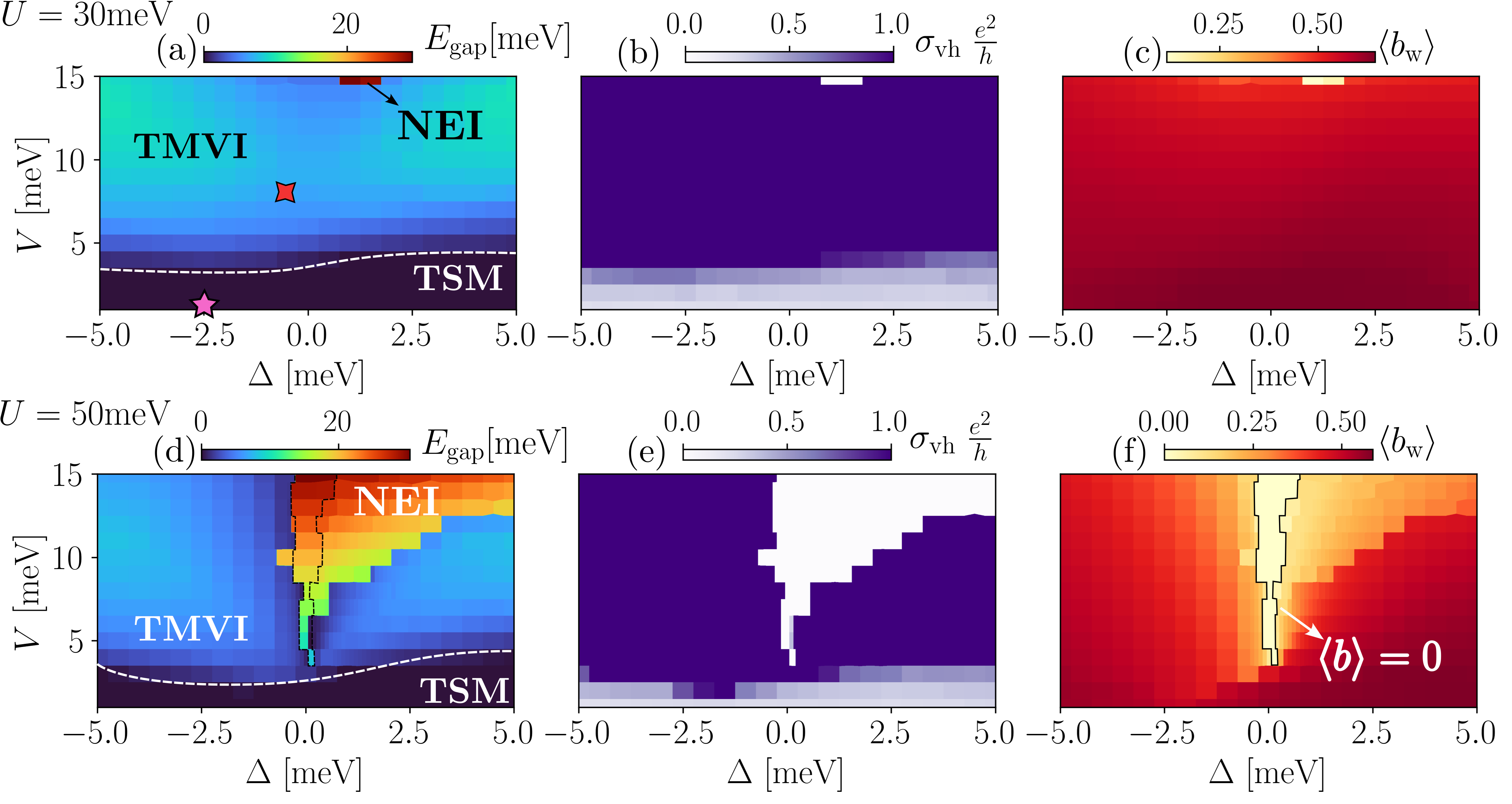} %0.9
\caption{Self-consistent parton mean-field phase diagram at $\nu_T=2$ as a function of
$\Delta$ and $V$ for a $12 \times 12$ system for (a)-(c) $U=30$ meV and (d)-(f) $U=50$ meV, respectively. The color codings denote the (a) and (d) charge-gap, (b) and (e) valley Hall conductivity, and (c) and (f) the boson condensate associated with the W holes. The dispersions and topology associated with the points marked in (a) are depicted in detail in Fig.~\ref{fig:effect_of_V} (a) and (b), respectively. 
}
\label{fig:mf_pd}
\end{figure*}

{\it Parton mean-field analysis.-} The numerical estimates of the energy scales places the system in the intermediate to strong-coupling regime. At filling $\nu_T=2$, when the individual layers are at filling $(1\pm\delta)$, a simple non-interacting or Hartree-Fock description might be insufficient to capture many of the essential features of the problem. Therefore, we implement a parton mean-field-based analysis, and obtain a theoretical phase diagram as a function of various microscopic tuning parameters. This will pave the way for a more sophisticated numerical analysis, which will also help guide future experiments. 

Consider the following representation for the electron operators, $c_{i,\sigma} = b_i f_{i,\sigma}$, where the  $b_i=e^{i\theta_i}$ represent spinless charged bosonic (``rotor") fields and the $f_{i,\sigma}$ denote spinful neutral spinons \cite{Florens_PRB_2004,Zhao_PRB_2007}. The parton representation entails an enlarged Hilbert space, which can be brought back to the physical Hilbert space by imposing the constraints $\langle n_i^\theta \rangle + \langle \sum_\sigma n^f_{i,\sigma}\rangle = 2$. Here, $n_i^\theta $ is the rotor charge and $n^f_{i,\sigma}$ is the fermionic occupation, respectively.
In what follows, $\sigma$ represents the conserved valley degree of freedom instead of physical spin. We begin by incorporating the mean-field decomposition of the above Hamiltonian in a reasonably conventional fashion, where the effect of the $U-$ term is included in the bosonic (rotor) sector. On the other hand, the effect of the $V-$ term is included in both the bosonic and fermionic sectors via variational weights, $\alpha$ and $(1-\alpha)$, respectively \cite{si}. 

The remaining variational parameters in the matter field sectors are the correlators, $B_{ij}\equiv \langle b^\dagger_i b^{\phantom\dagger}_j \rangle_\theta$ and $\chi_{ij,\sigma}\equiv \langle f_{i\sigma}^\dagger f^{\phantom\dagger}_{j\sigma}\rangle_f$, respectively. Note that in the absence of the $t_{\tn{hyb}}$ term (i.e. with enlarged $U(1)\times U(1)$ symmetry), a spontaneous expectation value for $\chi_{ij,\sigma}$ for nearest-neighbors generates an interlayer ``excitonic" order parameter. 
In order to solve the quartic bosonic Hamiltonian, we utilize a two-site cluster approximation for all pairs of $\langle ij \rangle$ and $\langle \langle ij \rangle \rangle$  \cite{Zhao_PRB_2007}; the self-consistent computations are carried out until the energy converges to within a small threshold \cite{si}. We also note that the numerical computations do not include the effects of emergent gauge-field fluctuations, but we comment on their effects at the end.

{\it Results.-} We have performed our mean-field analysis as a function of $\Delta$ and $V$ for two representative values of $U$.  
We restrict our attention to translation-invariant solutions, but allow for $C_3$ symmetry breaking. In order to characterize the different phases, we have used a variety of quantitative diagnostics, including the fermion gap ($E_{\tn{gap}}$), the fermionic valley Hall conductivity ($\sigma_{\tn{vh}}$), and the bosonic rotor condensate for the W holes ($\langle b_{\tn{W}}\rangle$). We note that, once the interlayer $B_{ij}$ develops, the bosonic condensates for the two layers share the same fate. The ``Mottness" associated with the suppression of charge fluctuations is characterized by the vanishing condensate. $\sigma_{\tn{vh}}$ is estimated by computing the Berry curvature of the renormalized fermionic band. According to the Ioffe-Larkin rule \cite{Ioffe_Larkin_PRB1989}, the electronic response is determined by the combination of the bosonic and fermionic responses \cite{Lee_Nagaosa_Wen}. Recall that when the bosons are condensed, the spinons become charged, and $E_{\tn{gap}}$ $(\sigma_{\tn{vh}})$ represents the electronic spectral-gap (conductivity). On the other hand, when the bosons are gapped, the system can potentially host fractionalized phases depending on the many-body state associated with the neutral spinons.

\begin{figure*}[pth!]
\centering
\includegraphics[width=1\linewidth]{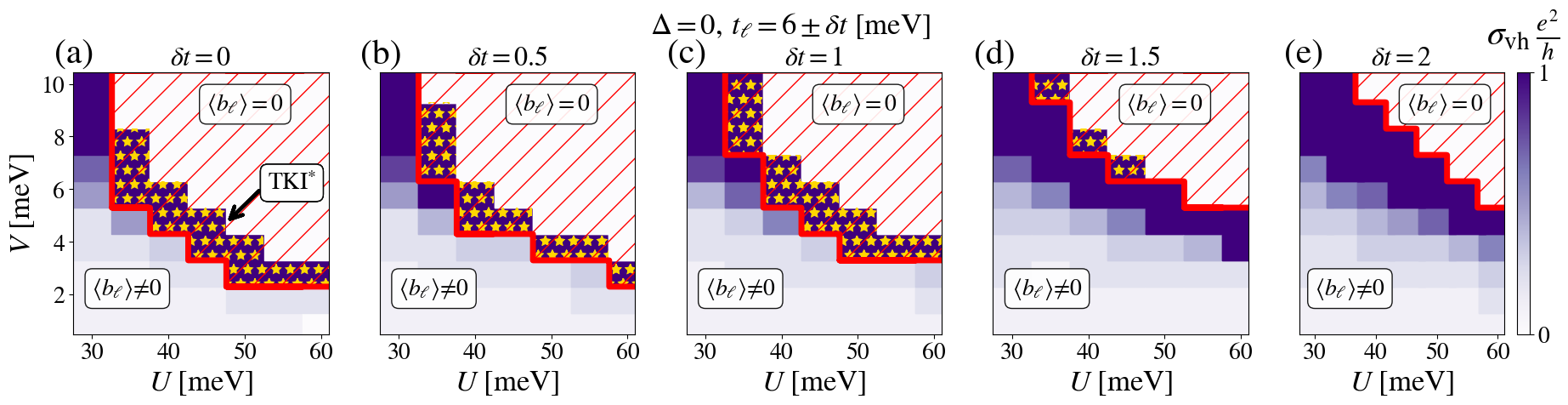}
\caption{Evolution of the fractionalized topological Kondo insulator (TKI$^*$; marked by $\star$) for various hopping anisotropies across the two layers, $t_\ell=6\pm\delta t$ (meV). The color-bar represents valley Hall conductivity $\sigma_{\tn{vh}}$. The regime with $\langle b_\ell \rangle\rightarrow 0$ appears to the right of the region demarcated by the red dashed lines; only the TKI$^*$ phase is stable against confinement.}
\label{fig:tki_fnof_delta}
\end{figure*}

Let us analyze the renormalized electronic dispersions and Berry curvature at two representative points in the phase diagram before a systematic scan over a wider range of parameters. Consider first the $V=0$ limit, as in Figs.~\ref{fig:effect_of_V} (a)-(b), for two different values of $\Delta$. We note that the bare bands (shown in gray) are strongly renormalized by the $U-$terms (colored bands in Fig.~\ref{fig:effect_of_V} a-b), which enter via the $\langle b_i\rangle,~\langle b^\dagger_i b^{\phantom\dagger}_j\rangle$ correlators in the parton analysis.  At zero displacement field, the band offset is larger than the bandwidth of both of the Mo and W bands, and the system is a trivial band insulator. With increasing displacement field, the WSe$_2$ band approaches the MoTe$_2$ band, leading eventually to a band-inversion. The $p-$wave inter-layer hybridization form-factor guarantees topological bands with a nonzero valley-contrasting Chern number. However, generically, the many-body state is {\it not} an insulator and is a compensated topological semimetal (TSM) instead, with $E_{\tn{gap}}=0$ and finite $\sigma_{\tn{vh}}$ over a broad range of $\Delta$. See Fig.~\ref{fig:effect_of_V} (a)-(b) for the fermiology associated with the compensated electron and hole pockets, as well as the momentum distribution of the Berry curvature associated with one of the valleys. It is worth emphasizing that in a purely {\it non-interacting} computation at $\nu_T=2$ with the bare bands and starting from the limit where $\nu_{\tn{Mo}}\approx 2$, the topological insulator survives in a rather small range of $\Delta$ \cite{si}.

Turning next to the effect of the nearest-neighbor repulsion, $V$, we find that it generically helps realize and stabilize insulating phases. In part, this is due to the enhanced excitonic tendency to bind the electrons and holes near the compensated limit for the two layers,  $\nu_\ell =1\pm \delta$ \cite{Excitonic_chern_dong_zhang_2023}. We find that for a critical $V\ll U$, a gap opening transition occurs (see Fig.~\ref{fig:mf_pd}(c)-(d)) and the resulting insulator has a quantized $\sigma_{\tn{vh}}$.  Note that the insulator also exists over a wide range of redistributed filling between the two layers (at the saddle-point level), which extends to $\langle \nu_{\tn{Mo}}\rangle$ away from $1$. We dub this state  topological mixed valence insulator (TMVI). As an insulating phase, the TMVI is smoothly connected to the non-interacting $\mathbb{Z}_2$ topological insulator discussed previously \cite{Zhao_qsh_2024}, but is crucially stabilized by the effects of interactions, which plays a number of different roles as we have seen above. To summarize, the main effect of $U$ is to drive band-flattening, which is more prominent for the MoTe$_2$ bands, and a small $V$ helps drive a Lifshitz transition from the TSM to the TMVI.

{\it Phase diagram.-} We have extended the analysis over a much wider range of parameters, as shown in the phase-diagrams for two different values of $U$; see Fig.~\ref{fig:mf_pd} (a), (d). The results in Fig.~\ref{fig:effect_of_V} for specific parameter sets are marked with the special symbols in Fig.~\ref{fig:mf_pd}(a), respectively. Both the TMVI and TSM host a non-trivial topological response, $\sigma_{\tn{vh}}\neq0$ (Fig.~\ref{fig:mf_pd} b, e) in a strongly renormalized electronic phase (i.e. without any fractionalization and $\langle b\rangle\neq0$; see Fig.~\ref{fig:mf_pd}c, f). 

We find an additional distinct phase with increasing $V$, driven by the competition between the TMVI and a topologically trivial ``excitonic" insulator; for $t_{\tn{hyb}}\neq0$ the excitonic character is \textit{not} sharply defined. Over a range of $\Delta$, the underlying $C_3$ symmetry is spontaneously broken and we dub the resulting state a nematic excitonic insulator (NEI) \cite{Xie_Pan_SDS_nematic_exciton}. The NEI is adiabatically connected to a dimerized excitonic insulator in the strong-coupling limit, $V\rightarrow \infty$. Although the interlayer excitonic order $\chi_{ij,\sigma}$ is finite for both the topological and trivial insulators, we reserve the term excitonic insulator specifically for the NEI, because the NEI develops along a specific bond an excitonic order much larger than that of the TMVI. We do not find evidence of any single-particle gap closing transition between the TMVI and the NEI, but the topological response $\sigma_{\tn{vh}}$ vanishes in the NEI. Other examples of a topological phase transition without single-particle gap closing have been discussed previously \cite{Ezawa_2013,Li_qah_2021,bradlyn}. Importantly, the transition above is accompanied by the disappearance of the nodal points of the form-factor, $\Phi_k$. The NEI spans over a broader extent in the phase diagram for larger $U$, as can be seen by comparing Figs.~\ref{fig:mf_pd} (a) and (d), respectively.

It is useful to comment briefly on the role of gauge-field fluctuations beyond the mean-field analysis at this point. Clearly, in the many-body phases with a condensed $\langle b_\ell\rangle\neq0$, and at $T=0$, the effects of the gauge-field fluctuations are irrelevant. On the other hand, notice that there is a range of values of $V,~\Delta$ in Fig.~\ref{fig:mf_pd}(f) where $\langle b\rangle =0$. Unsurprisingly, in this part of the phase diagram, the individual fillings associated with both layers self-consistently lock at $\langle \nu_\ell\rangle =1$, and the bosonic sector enters into a Mott-insulating phase. However, the spinons are fully gapped and are non-topological in this entire region; the sliver arises as a subset of the NEI phase. Therefore, in two spatial dimensions in the absence of any matter fields and non-trivial topology, the gauge-field will confine and the ground state will not exhibit any fractionalization \cite{polyakov}.

{\it Fractionalized Insulators.-} The appearance of a region with $\langle b\rangle=0$ and a nearly symmetric, self-organized ``Mott-like" phase raises an interesting possibility: Are there topologically non-trivial insulating phases with fractionalized excitations in the bulk? To address this question, we start from a theoretical limit of two identical layers, and perturb in a small parameter that controls their deviation away from this symmetric limit. For simplicity, we restrict our attention to $\Delta=0$. In the symmetric limit at total filling $\nu_T=2$ and large $U/t$, we are guaranteed to obtain two copies of a Mott insulator with the filling distributed as $\nu_T=1+1$. We then map out the many-body phase diagram as a function of the increasing hopping anisotropy $\delta t$, i.e. $t_\ell=6\pm\delta t$ (meV) over a range of $U$ and $V$ (Fig.~\ref{fig:tki_fnof_delta}). For $\delta t=0$, there is a wide range of parameters where $\langle b\rangle=0$, and the spinons remain gapped with a topological valley-hall response (region labeled by $\star$ in Fig.~\ref{fig:tki_fnof_delta}a). Note that the electronic response vanishes since the bosons are uncondensed. We dub this phase fractionalized topological Kondo insulator (TKI$^*$). With increasing $\delta t\lesssim 1.5$, there exists a finite regime in parameter space where the TKI$^*$ continues to remain the ground-state within parton mean-field theory (Fig.~\ref{fig:tki_fnof_delta}b-e). Note that the state is stable to inclusion of gauge-field fluctuations. For even larger $\delta t$, we find that the TKI$^*$ phase loses out the competition to the NEI; the local repulsion needed to gap out the bosons is large enough to flatten the fermionic bands and drive a nematic exitonic transition of the neutral fermions, which is unstable to confinement.

{\it Outlook.-} This work has been concerned with the emergence of novel correlated mixed-valence insulators with intertwined topological bands in TMD moir\'e heterobilayers. Over an extended regime in the many-body phase-diagram at total filling $\nu_T=2$, we find a strongly interaction-renormalized topological mixed-valence insulator with quantized valley Hall response, lying in close proximity to a topological semi-metal and trivial nematic ``excitonic" insulator. In the extended phase-diagram, we have also found the emergence of an interesting insulating TKI$^*$ phase with a fully gapped bulk and electrically neutral edge states. Finding a closely related tunable moir\'e system where this competition between different ground states can be studied experimentally remains a promising future direction. 

Looking ahead, the moir\'e bilayers provide an exciting playground for potentially realizing a number of closely related fractionalized phases. While the TKI$^*$ hosts a fully gapped bulk, the possibility of realizing a compensated topological semimetal of the electrically neutral spinons in the bulk (TSM$^*$) as an exotic gapless state, which is reminiscent of a candidate gapless ground-state for the three-dimensional mixed-valence insulators such as YbB$_{12}$ \cite{DCNC}, and quantum Hall bilayers \cite{Barkeshli_topo_neutral_exciton,Exciton2,Zibrov2017} remains an exciting prospect. While the parton-based methods here provide a key first step for exploring this rich phenomenology, future studies that rely on more sophisticated and numerically exact methods are needed to examine this physics using unbiased approaches, including the tendency towards additional forms of complex symmetry-breaking patterns \cite{tgk}.

%A distinction between our TMVI and typical mixed-valence (and Kondo) insulators lie in the fact that our TMVI arises from the hybridization within the same valley, rather than the same spin. The resulting excitons are thus spin-polarized. Therefore, the excitonic (and Kondo) criticality induced by an external magnetic field is expected to show a departure from the usual criticality in solid-state settings.

{\it Acknowledgements.-} We are grateful to Zhongdong Han, Kin Fai Mak and Jie Shan for numerous inspiring discussions related to their experimental data. We also thank Thomas Kiely and Dan Mao for a number of fruitful discussions. This work is funded in part by a Sloan research fellowship to DC from the Alfred P. Sloan foundation. 

{\it Note added.-} The day before this manuscript was being prepared for submission, an independent theoretical study analyzed related physics in moire heterobilayers \cite{guerci2024topologicalkondosemimetalinsulator}.

\bibliography{refs}

\clearpage

\begin{widetext}

%%%%%%%%%%
%\appendix 
\setcounter{page}{1} 
\setcounter{equation}{0} 
\renewcommand{\figurename}{Supplemental Figure}
\renewcommand{\theequation}{\thesection.\arabic{equation}}
%%%%%%%%%%%%%%%%

\begin{center}
    {\bf Supplementary material for ``Correlated Topological Mixed-Valence Insulators in  Moir\'e Heterobilayers"}\\
    Juan Felipe Mendez-Valderrama, Sunghoon Kim, Debanjan Chowdhury
\end{center}

\section{Description of the model}

\renewcommand{\thefigure}{S1}
\begin{figure}[h]
\includegraphics[width=0.5\linewidth]{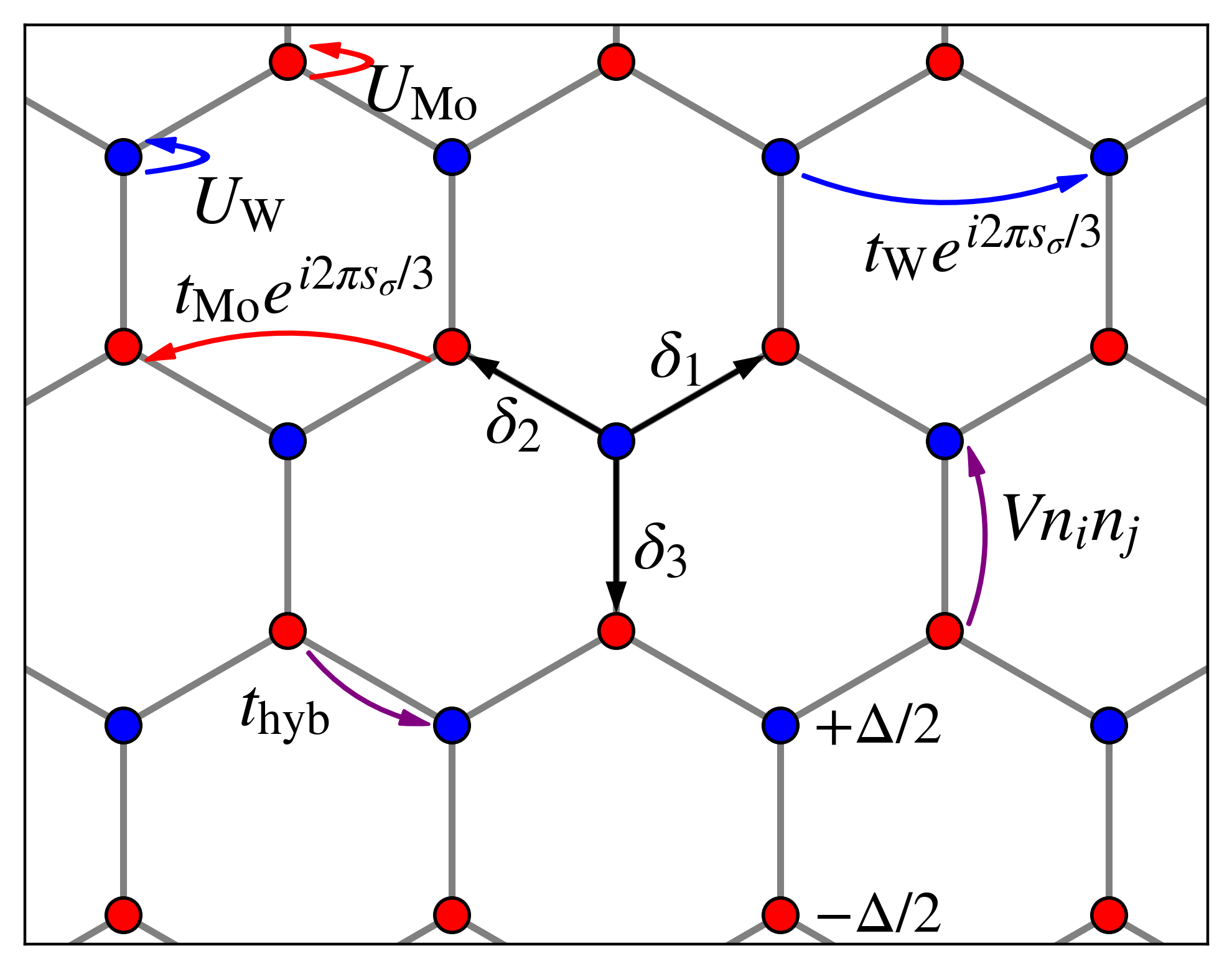}
\caption{Schematic of the interacting model defined on a honeycomb superlattice. Red (Mo) and blue (W) dots belong to two distinct sublattices of the honeycomb lattice. }
\label{fig:model}
\end{figure}

In Fig.~\ref{fig:model}, we present a schematic of the model studied in the main text, which illustrates all of the microscopic hopping and interaction parameters. 

\section{Topological phase transitions in the non-interacting limit}
\renewcommand{\thefigure}{S2}
\begin{figure}[h]
\includegraphics[width=1\linewidth]{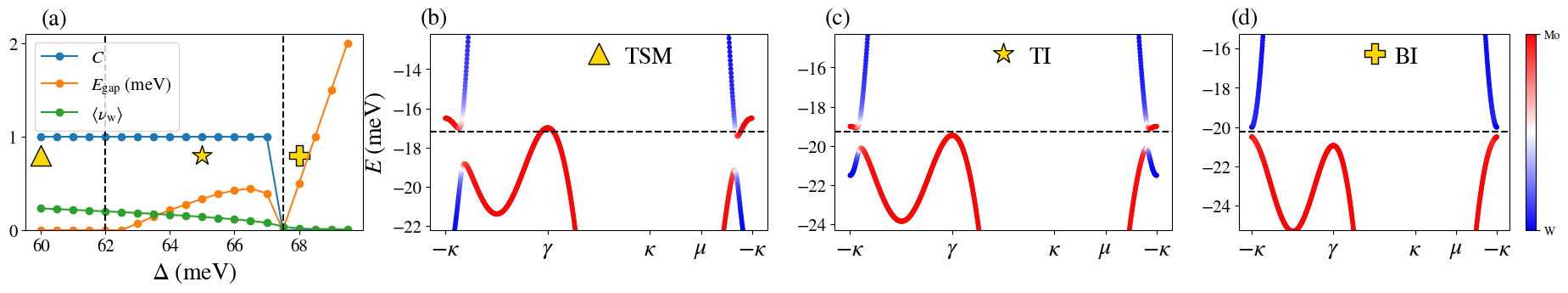}
\caption{Non-interacting phase diagram in the $(\nu_{\tn{Mo}},\nu_{\tn{W}})=(2-\delta,\delta)$ regime. (a) Evolution of $E_{\tn{gap}}$, the valley Chern number of the lower band, and $\langle \nu_{\tn W}\rangle$ as a function of $\Delta$. The black dashed lines mark phase boundaries. (b-d) Fermionic band dispersions for the 3 representative points shown in (a). The black dashed lines mark the Fermi levels. The dispersions and the Chern numbers are obtained for valley $K$.}
\label{fig:noninteracting}
\end{figure}
In this section, we present the phase diagram at $\nu_T=2$ in the non-interacting limit $(U=V=0)$. We focus on the $\nu_{\tn{Mo}}\approx 2$ regime near band inversion. At large $\Delta$, the system lies in a trivial band insulator (BI), where $\langle \nu_{\tn{Mo}} \rangle =2$. As $\Delta$ is decreased, the moir\'e gap decreases and eventually the band inversion takes place. The resulting phase is a $\mathbbm{Z}_2$ topological insulator (TI) with small gap. The TI is not as robust as in the interacting case, and immediately gives way to a topological semimetal (TSM) phase. As elaborated in the main text, strong $U$ and $V$ conspire to stabilize the gapped phase all the way to the mixed-valence regime.    

\section{Details of parton mean-field calculations}

In this section, we provide details of our parton mean-field calculations \cite{Florens_PRB_2004,Zhao_PRB_2007}. The mapping between the local Hilbert space of electrons and partons is given by 
\beq 
|0\rangle \equiv |2\rangle_\theta |0\rangle_f ,\nonumber\\
|\uparrow\rangle \equiv |1\rangle_\theta |\uparrow\rangle_f ,\nonumber\\
|\downarrow\rangle \equiv |1\rangle_\theta |\downarrow\rangle_f ,\nonumber\\
|\uparrow\downarrow\rangle \equiv |0\rangle_\theta |\uparrow\downarrow\rangle_f ,
\label{eq:parton_hilbert}
\eeq 
Here, ket states on the right hand side are direct product states of rotor and spinon ket states. The rotor ket states are labeled by rotor charge, $n^\theta$, and the spinon ket states are labeled by the (pseudo) spin degrees of freedom of the original electrons. For the parton representation, we impose the constraints $\langle n_i^\theta \rangle + \langle \sum_\sigma n^f_{i,\sigma}\rangle = 2$. 

As shown in the main text, the parton mean-field Hamiltonian is given by
\beq 
H&=&H^f +H^\theta, \nn \\
H^f&=&-\sum_{i\ne j,\sigma}t_{ij,\sigma}B_{ij}f^\dagger_{i\sigma}f_{j\sigma}-\mu \sum_{i}n_i^f  \pm \sum_i \frac{\Delta}{2} n_i^f , \nn\\
&-& (1-\alpha) V \sum_{\langle ij \rangle,\sigma} \chi_{ji,\sigma}f^\dagger_{i,\sigma}f_{j,\sigma},\nn \\
H^\theta &=&-\sum_{i\ne j}\sum_\sigma t_{ij,\sigma} \chi_{ij,\sigma}b^\dagger_{i}b_{j} + \frac{U}{2}\sum_i (2-n_i^\theta)(1-n_i^\theta)  \nn \\
&+&\alpha V\sum_{\langle ij \rangle} (2-n_i^\theta)(2-n_j^\theta) - \sum_i \mu_i^\theta n^\theta_i,
\eeq 
where $\alpha$, $B_{ij}\equiv \langle b^\dagger_i b_j \rangle_\theta$ and $\chi_{ij,\sigma}\equiv \langle f_{i\sigma}^\dagger f_{j\sigma}\rangle_f$ are variational parameters. The bosonic Hamiltonian is solved using two-site cluster approximation \cite{Zhao_PRB_2007}. To be specific, for every pair of sites $(i,j)$ with $|t_{ij,\sigma}|\ne 0$, we construct a cluster Hamiltonian. For instance, a cluster Hamiltonian for interlayer pair $\langle ij \rangle$ reads
\beq 
H^\theta_{ij} &=&-\sum_{a\in \{ij\}}\left(\sum_{k\notin \{ij\}}\sum_\sigma t_{ak,\sigma} \chi_{ak,\sigma}b^\dagger_{a}\langle b_{k} \rangle +\tn{h.c.} + \frac{U}{2} (2-n_a^\theta)(1-n_a^\theta)-  \mu_a^\theta n^\theta_a  \right)   \nn \\
&+& \alpha V (2-n_i^\theta)(2-n_j^\theta) - \sum_\sigma t_{ij,\sigma}\chi_{ij,\sigma}b^\dagger_i b_j +\tn{h.c.},
\eeq 
where $\langle b_k\rangle$ denotes the bosonic superfluid order parameter for the site $k$ outside of the cluster. Interlayer Coulomb terms for sites outside of the cluster are absorbed into the chemical potential term. For generic numerical parameters $(\{t_{ij,\sigma}\}, U, V, \Delta)$, we find that choosing $\alpha\rightarrow0$ minimizes the mean-field energy; see Fig.~\ref{fig:alpha_vs_Emf}. 

\renewcommand{\thefigure}{S3}
\begin{figure}[h]
\includegraphics[width=0.5\linewidth]{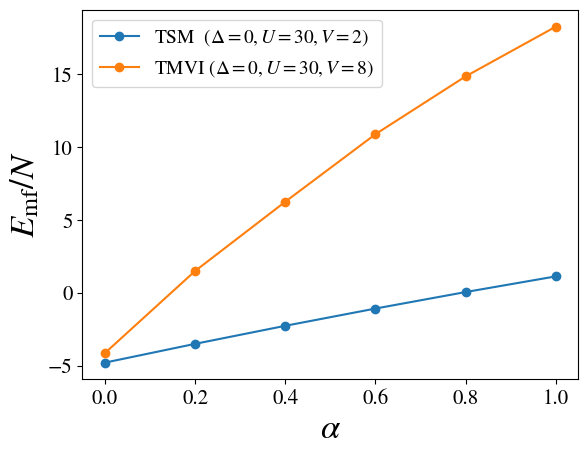}
\caption{Mean-field energy as a function of $\alpha$ for two representative points}
\label{fig:alpha_vs_Emf}
\end{figure}

Self-consistent calculations are performed in the following fashion. We focus on translationally-invariant solutions, while allowing for the $C_3$ symmetry breaking of correlators emanating from each site. We start by constructing initial $H^f$ with an ansatz for the fermionic correlator $\{\chi_{ij,\sigma}^{(0)}\}$. Using the initial fermionic ground state, we compute $\{\chi_{ij,\sigma}^{(1)}\}$ and the initial fillings $\langle \nu_{\tn{Mo}}\rangle=1+\delta$ and $\langle \nu_{\tn{W}}\rangle=1-\delta$. We then construct initial $H^\theta _{ij}$'s using $\{\chi_{ij,\sigma}^{(1)}\}$ and Lagrange multipliers $\{\mu_i^\theta\}$, which are chosen to fix the bosonic fillings to $\langle n_{\tn{Mo}}^\theta \rangle = 1-\delta$ and $\langle n_{\tn{W}}^\theta \rangle = 1+\delta$. The Lagrange multipliers can be obtained through an iterative procedure using the charge compressibility \cite{Hermele_PRB_chempot}. We compute bosonic expectation values $\{B_{ij}^{(1)}\},\{\langle b_i \rangle^{(1)}\}$ using the initial $H^\theta _{ij}$'s, where $\{\langle b_i \rangle^{(1)}\}$ is obtained by averaging over the results of all the cluster Hamiltonians involving the site $i$. We repeat this series of procedures to obtain a new set of variational parameters  
$\{\chi_{ij,\sigma}^{(2)}\},\{B_{ij}^{(2)}\},\{\langle b_i \rangle^{(2)}\}$. The self-consistent calculations are performed until the mean-field energy is converged within a threshold, i.e. $|E^{(n+1)}_{\tn{mf}}-E^{(n)}_{\tn{mf}}|/|E^{(n)}_{\tn{mf}}|<10^{-4}$. To obtain the mean-field phase diagram, we try $\sim 50$ random initial ansatz $\{\chi_{ij,\sigma}^{(0)}\}$', and choose the solution with a minimum mean-field energy.

\section{Phase diagram for different band parameters}

In Fig.~\ref{fig:mf_pdsi}, we present the mean-field phase diagram obtained for a slightly different set of band parameters with a larger anisotropy in the hopping parameters than the one for the DFT parameters: $(t_{\tn{Mo}},t_{\tn{W}},t_{\tn{hyb}},U)=(3,10,1,50)$ meV.

\renewcommand{\thefigure}{S4}
\begin{figure*}[pth!]
\centering
\includegraphics[width=1\linewidth]{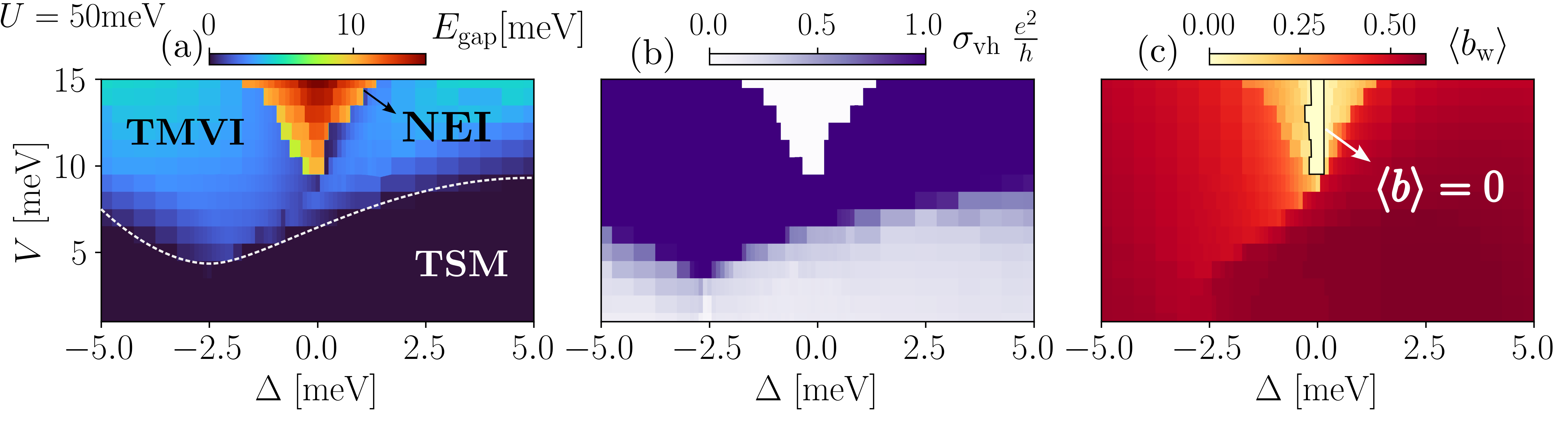} %0.9
\caption{ 
Mean-field phase diagram at $\nu_T=2$ as a function of
$\Delta$ and $V$ for system of size $12 \times 12$. Band parameters are chosen as $(t_{\tn{Mo}},t_{\tn{W}},t_{\tn{hyb}},U)=(3,10,1,50)$ meV. 
}
\label{fig:mf_pdsi}
\end{figure*}

\end{widetext}
\end{document}